\documentclass[journal,twoside,web]{ieeecolor}
\usepackage{tmi}
\usepackage{graphicx}
\usepackage{subfigure}
\usepackage{bm}
\usepackage{amsmath}
\usepackage{amssymb}
\usepackage{comment}
\usepackage{algorithm}
\usepackage{algorithmicx}
\usepackage{algpseudocode}
\usepackage{graphicx}
\usepackage{color}
\usepackage{url}
\usepackage{multirow}
\usepackage{booktabs}
\usepackage{multirow}
\usepackage{wrapfig}
\usepackage{enumerate}
\usepackage{caption}
\usepackage{cite}
\usepackage{bbding}
\usepackage{textcomp}

\usepackage[letterpaper=true,colorlinks,bookmarks=false]{hyperref}
\def\BibTeX{{\rm B\kern-.05em{\sc i\kern-.025em b}\kern-.08em
		T\kern-.1667em\lower.7ex\hbox{E}\kern-.125emX}}

\makeatletter

\newcommand{\Rmnum}[1]{\expandafter\@slowromancap\romannumeral #1@}
\definecolor{carminered}{rgb}{1.0, 0.0, 0.22}
\usepackage{stackengine}

\makeatother

\definecolor{light-gray}{gray}{0.8}

\makeatletter
\renewcommand{\maketag@@@}[1]{\hbox{\m@th\normalsize\normalfont#1}}%
\makeatother

\begin{document}

\title{Retinal Structure Detection in OCTA Image via Voting-based Multi-task Learning }

\author{Jinkui Hao, Ting Shen,  Xueli Zhu, Yonghuai Liu, Ardhendu Behera, Dan Zhang, Bang Chen, Jiang Liu, \\Jiong Zhang, Yitian Zhao
\thanks{This work was supported in part by the Zhejiang Provincial Natural Science Foundation of China (LR22F020008, LZ19F010001), in part by the Youth Innovation Promotion Association CAS (2021298), in part by the National Science Foundation Program of China (62103398), in part by the Ningbo Natural Science Foundation (202003N4039). (J. Hao and T. Shen contributed equally to this work.) (Corresponding authors:\textit{ \textbf{Xueli Zhu}}, shirleyzhu28@sina.com;\textit{ \textbf{Yitian Zhao}}, yitian.zhao@nimte.ac.cn)}
\thanks{J. Hao, B. Chen, J. Zhang, and Y. Zhao are with Cixi Institute of Biomedical Engineering, Ningbo Institute of Materials Technology and Engineering, Chinese Academy of Sciences, Ningbo, China; 
T. Shen is with Department of Ophtalmology, the Second Affiliated Hospital of Zhejiang University;
X. Zhu is with Ningbo First Hospital, Ningbo, China;
Y. Liu and A. Behera are with the Department of Computer Science, Edge Hill University, Ormskirk, UK.
D. Zhang is with School of Cyber Science and Engineering, Ningbo University of Technology, Ningbo, China
}
}
\maketitle

\begin{abstract}
Automated detection of retinal structures, such as retinal vessels (RV), the foveal avascular zone (FAZ), and retinal vascular junctions (RVJ), are of great importance for understanding diseases of the eye and clinical decision-making. In this paper, we propose a novel Voting-based Adaptive Feature Fusion multi-task network (VAFF-Net) for joint segmentation, detection, and classification of RV, FAZ, and RVJ in optical coherence tomography angiography (OCTA). A task-specific voting gate module is  proposed to adaptively extract and fuse different features for specific tasks at two levels: features at different spatial positions from a single encoder, and features from multiple encoders. In particular, since the complexity of the microvasculature in OCTA images makes simultaneous precise localization and classification of retinal vascular junctions into bifurcation/crossing a challenging task, we specifically design a task head by combining the heatmap regression and grid classification. We take advantage of three different \textit{en face} angiograms from various retinal layers, rather than following existing methods that use only a single \textit{en face}. 
To validate the superiority of our VAFF-Net, we carry out extensive experiments on three OCTA datasets acquired using different imaging devices, and the results demonstrate that the proposed method performs on the whole better than either the state-of-the-art single-purpose methods or existing multi-task learning solutions. We also demonstrate that our multi-task learning method generalizes across other imaging modalities, such as color fundus photography, and may potentially be used as a general multi-task learning tool. For the first time in the field of retinal OCTA image analysis, we construct three datasets for multiple structure detection. To facilitate further research, part of these datasets with the source code and evaluation benchmark have been released for public access.




\end{abstract}
 
\section{Introduction}

\label{sec-intro}
\begin{figure}[t]
    \centering{
    \includegraphics[width=1\linewidth]{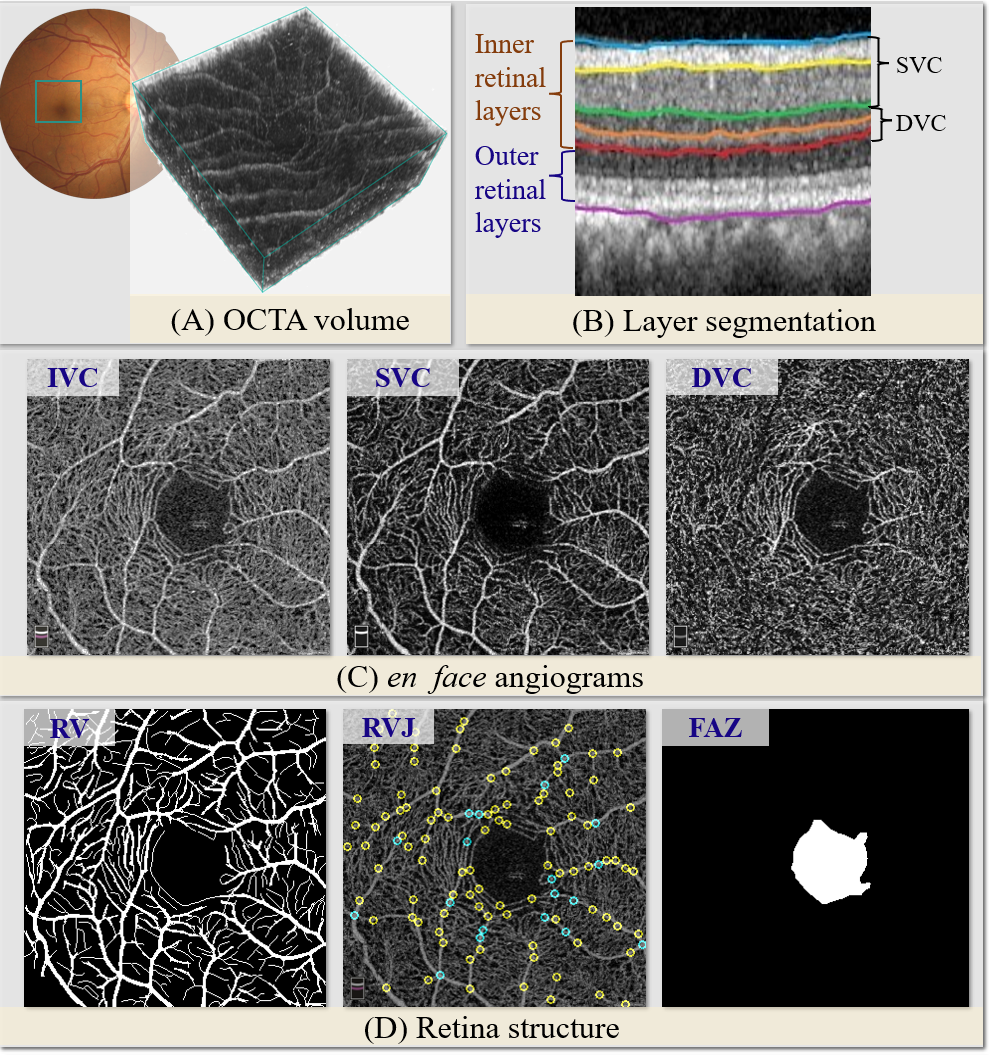}
    }
    \vspace{-10pt}
    \caption{Illustrations of the OCTA \textit{en face} angiograms and typical retina structure. (A) 3D OCTA volume. (B) Retinal layer segmentation of inner and outer retina. (C) The \textit{en face}  angiograms of the inner vascular complexes (IVC), superficial vascular complexes (SVC), and deep vascular complexes (DVC), respectively. (D) Typical retinal structures: RV, RVJ and FAZ.}
    \label{fig-octa}
\end{figure}

Optical coherence tomography angiography (OCTA) is a rapid and non-invasive imaging technique that can produce images containing functional information on retinal blood vessels and microvasculature ~\cite{kashani2017optical}.  Many studies have demonstrated that OCTA has advantages over other imaging modalities, in the detection and diagnosis of a variety of diseases of the eye ~\cite{Zang2020,RRobbins2020,cheng2021macular}. Compared to other retinal image modalities, such as color fundus images and  fluorescein angiography, OCTA can provide high-resolution 3D information about the retinal vasculature. Fig.~\ref{fig-octa} (A) illustrates a 3D OCTA volume within a 3$\times$3 $mm^2$ fovea-centered field of view. 

In general, the retina is composed of inner and outer retinal layers~\cite{campbell2017detailed}, as shown in Fig.~\ref{fig-octa} (B). By means of OCTA imaging technology, such as the RTVue XR Avanti SD-OCT system (Optovue, Fremont, USA), equipped with AngioVue software (version 2015.1.0.90), the inner vascular complexes (IVC) may be further subdivided into superficial vascular complexes (SVC) and deep vascular complexes (DVC):  their maximum projection of OCTA flow signals, a.k.a. \textit{en face} images, are shown in Fig.~\ref{fig-octa} (C). They allow for a clearer observation of the vasculature at different depth levels~\cite{campbell2017detailed}. 


Fig.~\ref{fig-octa} (D) demonstrates three typical retinal structures in OCTA images: retinal vessels (RV), foveal avascular zone (FAZ), and retinal vascular junctions (RVJ). (Note: cyan color indicates the vessel crossing  points,  and yellow indicates the vessel bifurcations.)
The quantification of these structures from the inner retinal layer by OCTA imaging plays a vital role in clinical decision-making of many diseases of the eye~\cite{spaide2018optical}.
For example, morphological changes in the FAZ are closely related to the conditions of age-related macular degeneration and glaucoma~\cite{cheng2021macular}. 
{It has been shown that the differentiation of bifurcations from crossings from the retinal vessel map is beneficial to the disease screening, diagnosis and progression understanding. For instance, the number of bifurcations was used as one biomarker to screen and diagnose Diabetic Retinopathy (DR)~\cite{sandhu2020automated,eladawi2018octa}.
In addition, the changes in angle of retinal vessel bifurcations was used in \cite{Akahori2018macular} to identify the macular displacement in OCTA.
On the other hand, distinguishing between bifurcation and crossing can also help 3D vessel reconstruction in OCTA~\cite{yu20213d}.
However, the existence of crossings usually interferes with the quantification of bifurcation-related features.
}
More recently, several studies have reported that the eyes of patients with Alzheimer’s disease show significantly reduced retinal vessel density when compared with healthy controls~\cite{RChua2020}, the demonstration of which demands accurate RV segmentation performance. To this end, an automated method to extract various retinal structures from OCTA has long been deemed desirable. 


Existing works for the extraction of retinal structures in OCTA images have largely adopted a single-task learning approach~\cite{diaz2019automatic,ma2020rose}. This approach usually designs a specific model for the detection of a single structure: occasionally, two models were integrated for detection of multiple structures. However, in these cases the integration scheme requires several individual training processes, which may lead to lower efficiency in terms of memory and computational requirements. A multi-task learning (MTL) approach provides a potential solution to address the above-mentioned issues.
MTL can perform multiple tasks simultaneously instead of establishing a set of independent networks. Compared with the single-task learning method, MTL not only allows for efficient extraction of multiple structures, but may also improve the quantification accuracy by exploiting the correlation, and complement of different tasks~\cite{Misra2016}. 
{Most of these approaches aim to handle multiple tasks from one model, the backbone needs to be carefully designed to achieve feature sharing. Moreover, these models often fail to consider the characteristics of different tasks and preferences for features.}
Meanwhile, current MTL works are usually designed for natural images~\cite{9336293,liu2020dynamic}, restricting their applicability to OCTA images due to the greater challenges in contrast, anatomical structure and imaging noise.

Most existing studies extract the retinal structure from an \textit{en face}  image of the IVC~\cite{peng2021fargo,liang2021foveal}. 
However, relying on a single 2D image as input does not make good use of the abundant sub-layer information provided by OCTA images~\cite{li2020image}. Consequently, we propose to use a multi-task learning approach to perform a joint detection of the RV, RVJ, and FAZ. This makes available SVC and DVC images as additional input data, as they can provide the depth-resolved information on the IVC. On one hand, the retinal vessels are mainly distributed in the SVC~\cite{an2020three}, and they can be extracted more accurately from the SVC since there is less projection noise. On the other hand, due to less interference by uncorrelated structures~\cite{li2020image}, the FAZ is easier to observe and extract in the DVC. It is worth noting that these different tasks share some similarities, so their joint learning will benefit each task individually. For example, due to the shared informative features, the RV detection may assist the identification of the RVJ, and a precise boundary of the RV around the macula contributes to the accurate  FAZ segmentation.

In order to further exploit the correlation and complement of the separate tasks, while taking into account the characteristics of each task, we use three encoders with shared weights to extract common feature representations from multiple inputs. Since each task has different characteristics and preferences for different inputs, we design a task-specific voting gate module to perform feature selection and fusion adaptively. This voting-based strategy allows for feature integration in both depth and plane dimensions based on the task attributes. In addition, we specifically design a task head by combining the heatmap regression and grid classification for the vascular bifurcation/crossing detection and classification tasks, as the complexity of the microvasculature in OCTA images makes it challenging for simultaneous precise localization and classification of the RVJ.

In summary, the key contributions of our work are:

$\bullet$~{We propose a voting-based adaptive feature fusion multi-task learning network for retinal structure detection in OCTA images, which can leverage the rich depth information of OCTA to obtain highly-accurate results. To our best knowledge, this is the first attempt to carry out the joint learning and detection of RV, FAZ, and RVJ within a single model with multiple inputs.}

$\bullet$~{As the most challenging sub-task, we specifically design a task head to differentiate the RVJ into bifurcations and crossings. This may potentially be used to address the problem of detection and classification of an unknown number of key points in complex backgrounds across images in different modalities.}

$\bullet$ {There are already some OCTA datasets available  for RV segmentation~\cite{ma2020rose,li2020ipn} or FAZ segmentation~\cite{agarwal2020foveal}, but there is no publicly available dataset for MTL and detection of RV, FAZ and RVJ in OCTA.} For the first time, we construct a publicly-accessible retinal structure detection dataset of OCTA images, with precise manual annotations of the RV, RVJ and FAZ. We also give a full evaluation/benchmarking of RV and FAZ segmentation, and RVJ detection and classification performance. The code of our method, comparative models, and evaluation tools are publicly available at \href{https://imed.nimte.ac.cn/ROSE-O.html}{https://imed.nimte.ac.cn/ROSE-O.html}.
\section{Related Works}
\label{sec-review}

We briefly review existing work on automated OCTA image analysis based on either single- or multi-task learning.
\vspace*{-0.2cm}
\subsection{Single-task based OCTA image analysis}

\subsubsection{RV segmentation}
The quantification of retinal vessels  plays a vital role in the study of eye diseases, and several specific segmentation methods have been proposed for detecting retinal vessels in OCTA images.
Eladawi et al.~\cite{eladawi2017automatic} presented a system for segmentation of retinal vessels from OCTA images based on a Markov-Gibbs random field model.
Mou et al.~\cite{mou2019cs} proposed a channel and spatial attention network (CS-Net) for curvilinear structure segmentation, including vessel segmentation in OCTA images. 
Li et al.~\cite{li2020image} introduced a 3D-to-2D image projection network to achieve the 2D retinal vessel segmentation from 3D volume data. (It is worth noting that they also performed the FAZ segmentation using this method, but the two tasks were implemented individually rather than through joint learning.)
Giarratano et al.~\cite{giarratano2020automated} first applied handcrafted filters and neural network architectures to enhance the vessels, and obtained the final vessel segmentation results using deep learning methods. Recently, Ma et al.~\cite{ma2020rose} proposed a coarse-to-fine network with split attention to segment vessels, and evaluated it on the public ROSE dataset.

\subsubsection{RVJ detection and classification}
The number of bifurcations in OCTA images can be used to diagnose diabetic retinopathy~\cite{sandhu2020automated,eladawi2018octa}. However, current RVJ detection and classification studies are all based on color fundus photography (CFP)~\cite{morales2017retinal,calvo2011automatic,Zhao2020VJ}.
For example, a feature point detection and classification method based on local and topological analysis was proposed in~\cite{calvo2011automatic} for CFP images.
Morales et al.~\cite{morales2017retinal} first determined the vessel skeletons using a stochastic watershed transformation, and the junction points were detected by template matching and then classified into crossing or bifurcation using closed loop check.
Zhao et al.~\cite{Zhao2020VJ} proposed a method using deep neural networks to detect and classify the retinal vascular junctions in CFP images.
The detection and classification of bifurcations and crossings in OCTA images is under-studied, due to various challenges such as high vessel complexity and low capillary visibility.

\subsubsection{FAZ segmentation}
Many studies have showed that the shape alteration of the FAZ is related to the onset and progression of diabetic retinopathy~\cite{rosen2019earliest},  so a method for FAZ segmentation in OCTA is desired. 
Díaz et al.~\cite{diaz2019automatic} proposed a fully automated system by using a series of morphological operators to identify and precisely segment the region of the FAZ.
Guo et al.~\cite{guo2019automatic} introduced  a deep learning network with an encoder–decoder architecture for FAZ segmentation from OCTA.
Li et al.~\cite{li2020fast} presented a lightweight U-Net to segment the FAZ from OCT and OCTA projection maps.

\begin{figure*}[t]
    \centering{
    \includegraphics[width=0.95\linewidth]{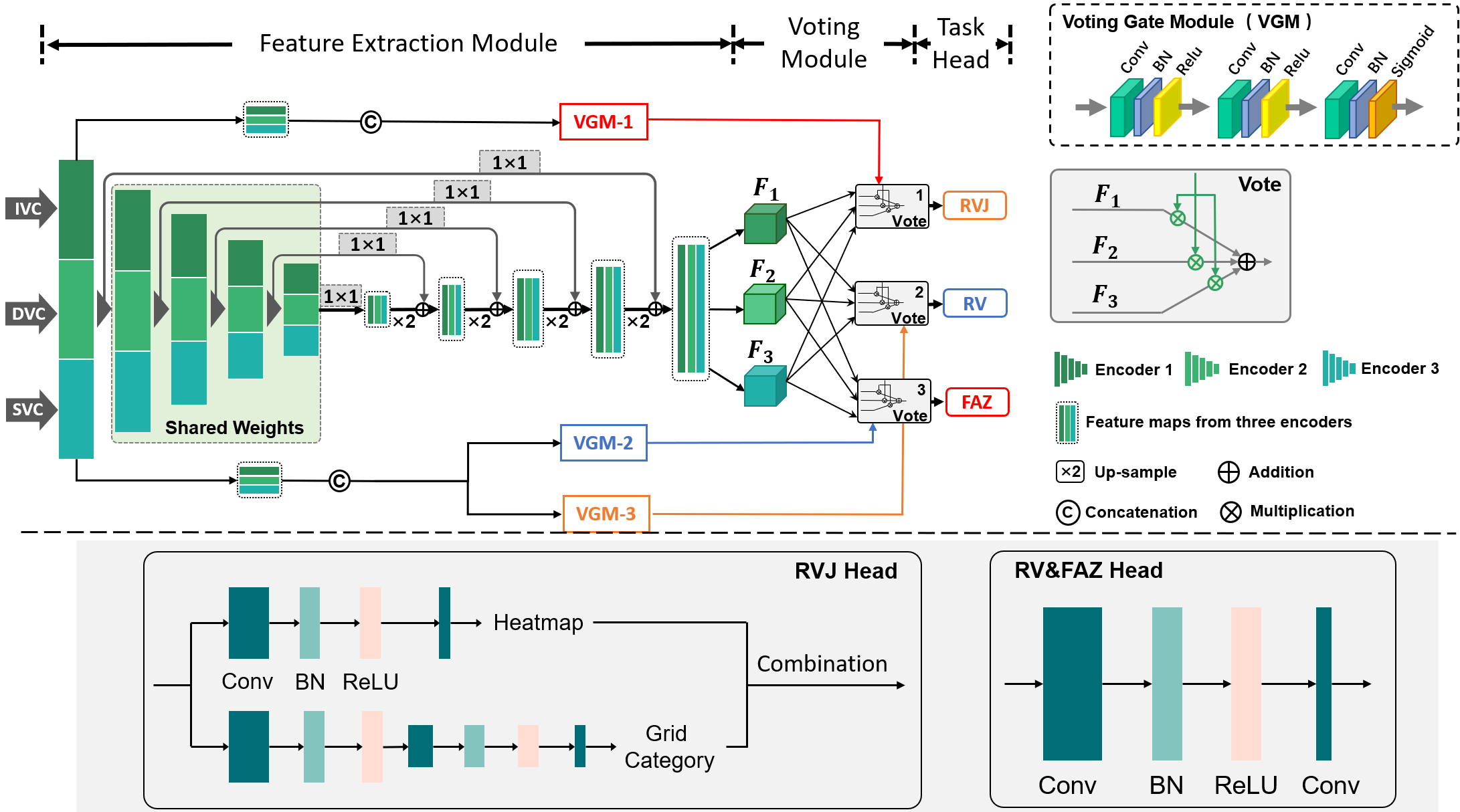}
    }
    \caption{The architecture of our VAFF-Net, consisting of the feature extraction module, the voting module and the task head. }
    \label{fig-network}
    \vspace{-12pt}
\end{figure*}

\subsection{Multi-task learning and its applications in OCTA}
Multi-task learning has been broadly used in computer vision, aiming to improve learning efficiency and prediction accuracy for multiple tasks~\cite{Misra2016,liu2019end,9336293}.
Most MTL methods focus on two aspects: the design of network architectures~\cite{Misra2016,ahn2019deep}; and loss functions to balance the importance of different tasks~\cite{kendall2018multi}.
CrossStitch Networks~\cite{Misra2016} utilized cross-stitch units to combine multi-task neural activations and allow features to be shared across tasks.
MTAN~\cite{Liu2019} used a shared backbone network in conjunction with task-specific attention modules in the encoder to learn task-specific features.
PAD-Net~\cite{xu2018pad} introduced a decoder-focused architecture, in which the backbone features are processed by a set of task-specific heads to produce the prediction for every task.
Kendall et al.~\cite{kendall2018multi} used homoscedastic uncertainty to balance the single-task losses.
We refer readers to~\cite{9336293} for the latest and comprehensive comparisons between different MTL approaches. The majority of these methods focus on natural images, taking individual images as input, which restrict their applicability to medical images with limited contrast and complex and varied structures and distributions, such as OCTA images.
{The success of MTL over natural images has been unsurprisingly extended to the medical imaging applications~\cite{de2018clinically,guo2021end}. For example, Liu et al. ~\cite{liu2018joint} proposed a MTL framework for simultaneous brain disease classification and clinical score calculation. 
Gao et al.~\cite{gao2020feature} introduced a feature transfer enabled MTL method for joint detection, segmentation and classification for breast cancer diagnosis.
He et al.~\cite{he2020multi} proposed a method for joint learning for two tasks in parallel in the CT images: the segmentation and multi-label classification of organs.
The study from Amyar et al~\cite{amyar2020multi} introduced a MTL model to jointly identify COVID-19 patients and segment COVID-19 lesions from chest CT images. These methods all use a single input, and most of them consist of a main shared deep CNN architecture and multiple different task heads, without considering the preferences of different tasks for features.}

Recently, several MTL methods for OCTA image analysis have been proposed.
Peng et al.~\cite{peng2021fargo} proposed a framework for simultaneous segmentation of the FAZ and RV from OCTA images. They utilized spatial and channel attention modules to improve segmentation performance.
Lin et al.~\cite{lin2021bsda} introduced a joint learning method for FAZ segmentation and diagnostic classification. They used the detected FAZ to improve the performance of diagnostic classification networks. 
Wang et al.~\cite{wang2021diagnosing} proposed an MTL method for simultaneous diagnosis and segmentation of choroidal neovascularization in OCTA images.
However, these works are direct applications of existing MTL methods, and do not exploit the unique characteristics of OCTA images, such as rich depth information.

\section{Datasets}
\label{sec:data}

We released an OCTA image dataset for vessel segmentation, namely ROSE (\textbf{R}etinal \textbf{O}CTA \textbf{SE}gmentation), along side with our previous work~\cite{ma2020rose}. In order to train a more robust and widely-applicable retinal structure detection model, for this work we further constructed three data subsets ROSE-O, ROSE-Z, and ROSE-H for the RV, FAZ and RVJ detection and classification, respectively. The images were acquired using the three most commonly-used OCT imaging devices worldwide: \textbf{ \underline{O}}ptovue Avanti RTVue XR (Optovue, Fremont, USA), \textbf{\underline{Z}}eiss Cirrus HD-OCT 5000 (Zeiss Meditec, Dublin, USA), and \textbf{ \underline{H}}eidelberg Spectralis OCT2 (Heidelberg Engineering, Heidelberg, Germany).  All the data were collected under the approvals of relevant authorities and consent of the patients, following the Declaration of Helsinki. 

$\bullet$ \textbf{{ROSE-O}} contains 117 images which were captured using the Optovue Avanti RTVue XR with AngioVue software (Optovue, Fremont, USA): the images have a resolution of $304 \times 304$ pixels. The SVC, DVC and IVC angiograms of each participant were obtained using the device. {In order to ensure a fair comparison, we followed the setting in ~\cite{ma2020rose} when dividing the dataset for training and testing,  i.e., the dataset was split into 90 images for training and 27 images for testing.}
{The training set contains 20 AD subjects and 10 healthy controls and the test set contains 6 AD subjects and 3 healthy controls.}

$\bullet$ \textbf{{ROSE-Z}} was acquired using Zeiss Cirrus HD-OCT 5000 with AngioPlex software (Zeiss Meditec, Dublin, USA). It consists of 126 OCTA images from 42 subjects (15 with diabetic retinopathy, 2 with Alzheimer’s disease and 25 healthy controls), and each subject had their SVC, DVC and IVC \textit{en face}  scans. All \textit{en face} images cover a field of view of $3 \times 3mm^2$ centered at the fovea with $512 \times 512$ pixels. The dataset was randomly split at subject level. 
{
5-fold cross-validation was used to evaluate the models on this relative small dataset.}


$\bullet$ \textbf{{ROSE-H}} consists of 60 OCTA images from 20 eyes (including 8 with choroidal neovascularization and 12 healthy controls) acquired using the Heidelberg Spectralis OCT2 (Heidelberg Engineering, Germany).  All \textit{en face} images cover a field of view of $3 \times 3mm^2$  centered at the fovea with $304 \times 304$ pixels. Each subject had their SVC, DVC and IVC \textit{en face}. {All the images in the ROSE-H dataset were used for testing only in our experiments to assess the generalizability of different methods over images captured using a device from another manufacturer.}

For all the datasets, three well-trained imaging experts manually labeled the RV, FAZ, and RVJ (including the differentiation between bifurcations and crossings) in the IVC images, then two senior ophthalmologists with more than 15 years’ clinical experience reviewed and refined the annotations. Their consensus was finally defined as ground truth for the purpose of this study. 
{
FAZ is the fovea devoid of capillaries in the macula that can be described as a dark area without vessels at its center~\cite{eladawi2018early}. During the annotation process, we followed the following criteria:  FAZ is defined as the largest closed loop surrounded by capillaries in the fovea of the macula in OCTA. The FAZ should contain as few high-intensity signals as possible, and its boundaries should be as close as possible to the surrounding capillary plexus. 
According to the previous study~\cite{nesper2017quantifying}, the blood vessels were defined as pixels having decorrelation values above the threshold level of noise.  Therefore, we labeled the pixels with significantly higher pixel values than the FAZ area as blood vessels. Meanwhile, since the vessels and capillaries are continuous structures, according to the suggestion of the ophthalmologists, we treated the isolated points as noise and then excluded them.
Based on the previous study~\cite{dashtbozorg2013automatic}, the classification of the intersections of the vessel map may be broken down into two categories: (i) the bifurcation – different blood vessel segments are from one blood vessel tree, and (ii) the crossing – two blood vessels overlap due to the projection of a 3D human eye to a 2D \textit{en face} image.
The vessels separated in 3D space may intersect in the \textit{en face} image obtained through maximum projection. Therefore, if four or more vessel segments meet at a point, it is considered a crossing. 
}
Fig.~\ref{fig-octa} (D) illustrates sample manual annotations of RV, RVJ, and FAZ respectively. ROSE-O has been released for public access. 
~{Note that ROSE-O and ROSE-1 released in ~\cite{ma2020rose} have the same samples. Their difference lies in that in addition to the annotation of vessel segments, ROSE-O also contains the new annotations of FAZ and bifurcations/crossings. ROSE-Z and ROSE-H are  newly collected datasets.}



\vspace*{-0.2cm}
\section{Proposed method} 
\label{sec:method}
In this section, we detail our proposed VAFF-Net, including its architecture, a specific task head for detection and classification of RVJ, and the loss function for its end-to-end training. 
\vspace*{-0.2cm}
\subsection{Architecture}
\label{sec-extractor}
The overall architecture of the proposed network is illustrated in Fig.~\ref{fig-network}. Our VAFF-Net includes three main components: the feature extraction module, the voting gate module (VGM), and the task head. 
{The purpose of our model is to extract multiple retinal structures simultaneously using inputs containing depth information. The input of VAFF-Net is three enface projections including IVC, SVC and DVC. Through the feature extractor and three task heads, we can obtain RV, FAZ and RVJ detection results simultaneously.}
The feature extraction module consists of three feature extractors, which correspond to the three input \textit{en face}  angiograms, i.e., IVC, SVC and DVC.
We apply ResNet-50~\cite{he2016deep} as feature extractors, in which the first $7\times7$ convolutional layer is replaced by a $3\times3$ convolution with the same padding to ensure that the output size of the voting gate module is consistent with the size of the input image. In our implementation,  three extractors share weights except the first convolutional layer, in order to limit the number of learnable parameters. 
Due to the different inputs and the independence of the first layer, these three encoders are able to extract different features, despite our strategy in sharing weights later.

The voting module contains three independent task-specific voting gate modules, each corresponding to a task and adaptively learning how to perform feature selection and fusion.
The VGM consists of multiple $3\times3$ convolutional layers with batch normalization (BN) and ReLU activation, the final convolutional layer with a sigmoid operator is used to map the features into the form of probability with 3 channels that can be utilized as the gate to select features.
The VGM for each $task \in \left\{RV, FAZ, RVJ\right\}$ takes the concatenation of the output of the first layer from the three encoders as input, and the corresponding output $\left\{G_{task}\right\}$ is the learned voting gate, which can select features at two levels: features from different layer slabs \textit{en face} images, and features at different spatial locations from an encoder. The former takes into consideration that the importance of the features obtained from the three $en face$ images is different for each task. For the latter, we may utilize the following spatial characteristics: the FAZ segmentation task focuses on the macular area; the RVJ detection task relies on the location of the intersecting vessels, and the vessel segmentation task requires greater attention to the edges of the vessels. 

After having obtained the voting gate $\left\{G_{task}\right\}$ for each task, the multi-scale fused features $\left\{F_{i}\right\}(i \in \left\{1,2,3\right\})$ from three encoders are multiplied with $\left\{G_{task}\right\}(task \in \left\{RV, FAZ, RVJ\right\})$ respectively, and a summation  is performed to obtain the integrated feature map $\left\{M_{task}\right\}$ for the corresponding task. These operations can be formulated as:
\begin{equation}\small
    \label{eq1}
    M_{task}=\sum_{i=1}^{n}{G_{task}^i\circ F_i},
\end{equation}
where  $n$ is the number of feature channels, $\left\{G_{task}^i\right\}$ indicates the $i${th} channel of voting gate $\left\{G_{task}\right\}$, and $`\circ$' denotes the element-wise multiplication. The task-specific feature map $\left\{M_{task}\right\}$ is then fed into the corresponding task head to obtain the final task-specific result.

\subsection{Task head for RVJ detection and classification}
The task for detection of keypoints on human body in computer vision usually involves detecting a number of keypoints that is known in advance~\cite{newell2016stacked}; but the number of RVJs varies from one subject to another. Furthermore, the RVJs are small targets covering only a few pixels, and  the bounding box-based approaches for object detection~\cite{ren2016faster, duan2019centernet} usually have difficulty in achieving satisfactory performance on the RVJ detection task. In consequence, we consider that the RVJ detection and classification impose a greater challenge to this multi-task learning framework.  

\begin{figure}[h]
    \centering{
    \includegraphics[width=1\linewidth]{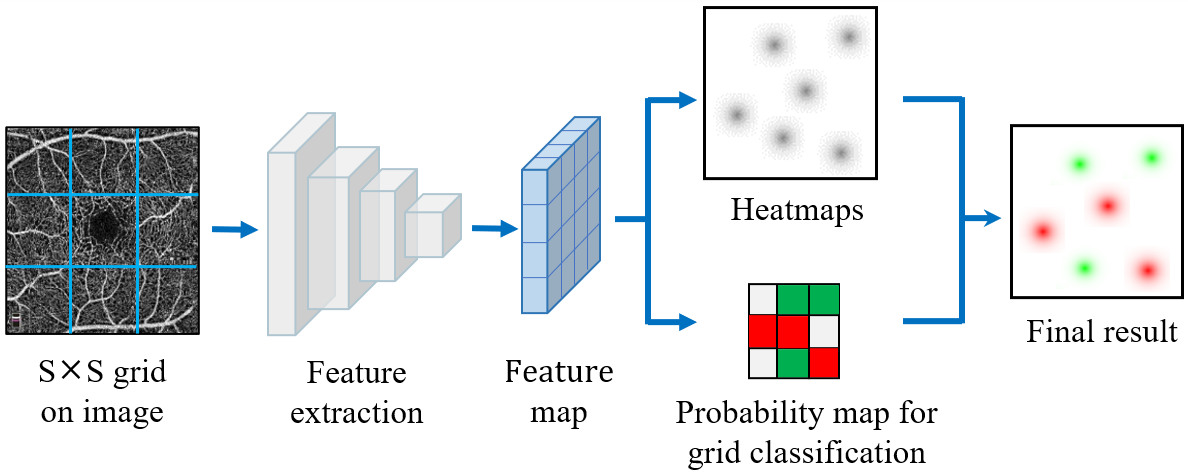}
    }
    \vspace{-6pt}
    \caption{Overview of the task head for RVJ detection and classification. }
    \label{fig-junchead}
    \vspace{-10pt}
\end{figure}

In order to address these issues, we introduce a task head with two branches, that combines heatmap regression and grid classification for the detection and classification of bifurcations and crossings.
We split this relatively complex task into two simple ones: heatmap regression is utilized to locate all the RVJs, and the grid classification branch is used to distinguish the RVJs between bifurcations and crossings. Fig.~\ref{fig-junchead} shows the architecture of the proposed RVJ detection and classification head. 

{The input of the RVJ head is the output of the feature extractor reweighted by VGM of RVJ, i.e., $\left\{M_{RVJ}\right\}$ mentioned in Sec.~\ref{sec-extractor}.}
The feature map $\left\{M_{RVJ}\right\}$ for RVJ detection is first fed into a convolutional block, which consists of two $3\times3$ convolutional layers with BN and ReLU activation functions. 
The last convolutional layer with the sigmoid activation function obtains the locations of all the junctions through the heatmap output of 1 channel.
Another branch also takes  $\left\{M_{RVJ}\right\}$ as input: it divides the image into an $S\times S$ grid, and for each grid cell predicts 3 class probabilities (i.e., containing bifurcation, containing crossing, and background only) and 1 confidence score. These predicted values are encoded into an $S\times S\times 4$ tensor.
The confidence score indicates how confident the model is that the grid contains an RVJ, and can be used to select a threshold during final processing. 
As the model focuses on the prediction of grid types only, and does not need to predict the coordinates of the bounding box, we believe that our method may achieve better performance compared to the bounding box-based one~\cite{Zhao2020VJ}, as evidenced in Sec. VI.D. In our implementation, we set each grid cell as $8\times8$, and the final prediction of this branch is thus a $38\times38\times4$ tensor for an image with an input of $304\times304$. 
{The output of one branch is a heatmap of all the junctions, and the output of the other branch is the category of the junction contained in each grid.}
We obtain the final predictions by combining the results of the two branches. 
~{The size of the grid is a hyperparameter. For the input of 304x304, we empirically found that the grid size of 8x8 is appropriate, so as to ensure that there is at most one bifurcation/intersection in a grid as much as possible. The size of the grid can be adjusted according to the size of the input image.}



\subsection{Loss function}
For an \textit{N}-task problem with task losses $\mathcal{L}_1$, $\mathcal{L}_2$, ..., and $\mathcal{L}_N$, the total loss function is expressed as:
\begin{equation} \small
\mathcal{L}_{Total}=\sum_{n=1}^N\lambda_n\mathcal{L}_n,
\end{equation}
where $\lambda_n$ is the weight of task-specific loss $\mathcal{L}_n$. We use the standard binary cross-entropy (BCE) loss for both the RV and FAZ segmentation tasks. Given a prediction map $\hat{Y}$ and corresponding ground truth $Y$, the standard binary cross-entropy loss is $\mathcal{L}_{BCE}(\hat{Y},Y)$. Therefore, the training losses for RV and FAZ segmentations are $\mathcal{L}_{BCE}(\hat{Y}_{RV},Y_{RV})$ and $\mathcal{L}_{BCE}(\hat{Y}_{FAZ},Y_{FAZ})$, respectively.


The training loss in the RVJ detection task consists of heatmap prediction loss and grid classification loss.
We generate the ground truth for heatmap regression and grid classification based on the annotated pixel coordinates of the RVJs, where the heatmap is generated using a Gaussian kernel with a standard deviation of 2.5.
The heatmap regression branch is trained using the mean squared error (MSE) loss: $\mathcal{L}_{MSE}(\hat{Y}_{map},Y_{map})$. 
For the grid classification branch, the training loss includes classification and confidence score errors. Since many grid cells do not contain any junction objects, this would push the confidence scores of those cells towards zero, thus overpowering the gradients from the cells that contain junctions. This leads to model instability, causing the training process to diverge early~\cite{Redmon2016}. To this end, we use the hyperparameters $\lambda_{A}$ and $\lambda_{B}$ to balance the loss of grids containing and not containing junction objects.
The loss function for the grid classification branch is thus as follows:
\begin{equation} \small
    \begin{split}
        \mathcal{L}(\hat{Y}_{grid},Y_{grid})=&\lambda_{A}\sum_{i=1}^{S^2}\natural_{i}^{A}(C_i-\hat{C_i})^2\\
        &+\lambda_{B}\sum_{i=1}^{S^2}\natural_{i}^{B}(C_i-\hat{C_i})^2\\
        &+\sum_{i=1}^{S^2}\sum_{c\in{classes}}(p_i(c)-\hat{p}_i(c))^2,
    \end{split}
\end{equation}
where $\natural_{i}^{A}$ denotes if the junction object appears in grid $i$, and $\natural_{i}^{B}$  otherwise. $C_i$ and $\hat{C_i}$ denote the ground truth and prediction confidence scores of grid $i$, respectively. $p_i(c)$ and $\hat{p}_i(c)$ denote the ground truth and prediction probability of class $c$ for grid $i$. The hyperparameters $\lambda_{A}$ and $\lambda_{B}$ are empirically set as 5 and 1, respectively. The performance of the multi-task learning model is heavily dependent on the relative weighting between the losses in different tasks. In our implementation, we selected the Dynamic Weight Average (DWA)~\cite{liu2019end} for all the multi-task learning methods, and the default setting was employed to train our model.




    \vspace*{-0.3cm}    
\section{Experimental results}
In this section, we describe the implementation details, evaluation metrics, and experimental results.




   \vspace*{-0.3cm}    
\subsection{Implementation details}
The proposed network was implemented using Python in the PyTorch package. All experiments were carried out on a workstation containing two NVIDIA GeForce GTX 3090 with a memory of 24GB. The Adam optimizer with recommended parameters was used to optimize the model, and the batch size was set as 4.~{The total number of epochs was set to 1000 for network training. The initial learning rate was $5\times 10^{-5}$ and gradually decayed to zero after 1000 epochs using a Cosine annealing scheduler. 
Data augmentations were conducted during all the training stages, including random horizontal flip, vertical flip, rotation of the image by an angle from $-10^{\circ}$  to $10^{\circ}$ around its center, and gamma transformation with a range of (0.7, 1.9). 
All the images were normalized from [0, 255] to [0, 1] before being fed into the model.}
   \vspace*{-0.2cm}    
\subsection{Evaluation metrics}
We use three metrics to evaluate the RV and FAZ segmentation performances:
\begin{itemize}
\item {Dice coefficient (DICE) = 2 $\times$ TP / (2 $\times$ TP + FP + FN);}
\item {{Balance-Accuracy (BACC) = ((TP / (TP + FN)) + (TN / (TN + FP)) / 2;}}
\end{itemize}
For the detection and classification of RVJs, the following metrics are calculated:
\begin{itemize}
\item {Recall (RE) = TP / (TP + FN)}
\item {F1-score = (2 $\times$ PR $\times$ RE) / (PR + RE)}
\end{itemize}
where TP, FP, TN, and FN denote true positive, false positive, true negative, and false negative. TPR is the true positive rate, and TNR is the true negative rate. 
{
Our method locates the junctions at single-pixel level, but considering that the location of a junction is not restricted to a single pixel, and the width of the blood vessel intersection in OCTA images is usually more than 5 pixels, 
the tolerance was also set to 5 following the evaluation setting of previous studies
~\cite{Zhao2020VJ,Hervella2020}.
}

\begin{figure*}[t]
    \centering{
    \includegraphics[width=0.9\linewidth]{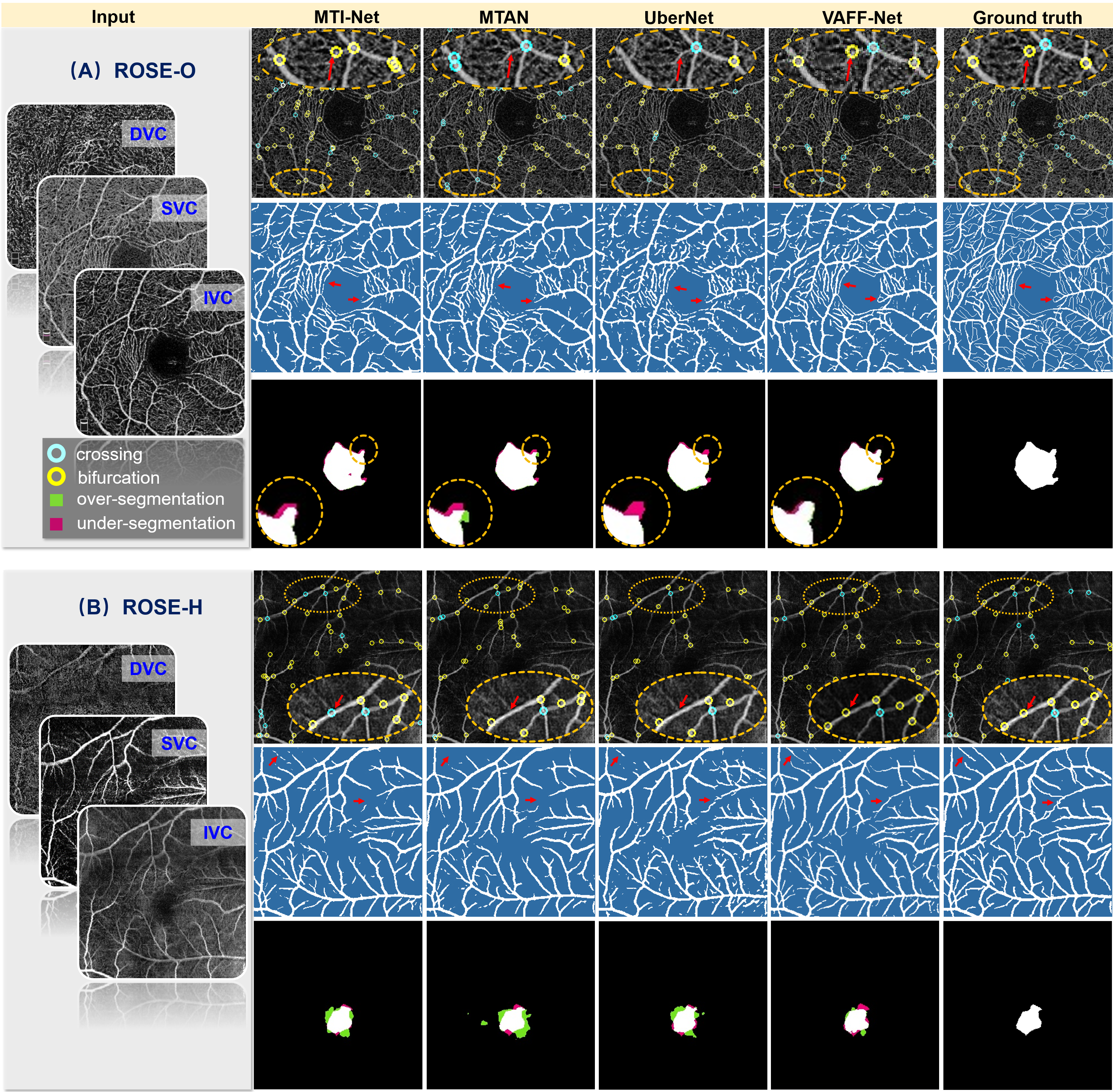}
    }
    \caption{{The detection results of different methods on a randomly selected case from \textbf{ROSE-O} and \textbf{ROSE-H} respectively. Top to bottom for one case: the results of RVJ detection and classification, RV segmentation, and FAZ segmentation, respectively.} 
    }
    \label{fig-resOCTA}
    \vspace{-10pt}
\end{figure*}

\vspace*{-0.2cm}    
\subsection{Results}

We compare the proposed method with the state-of-the-art ones on RV and FAZ segmentation, and RVJ detection and classification, for the single-purpose methods, and the multi-task learning approaches over all the three tasks. It is important to note  that since our model requires SVC, DVC and IVC as input, for fair comparison we concatenated these three images as input for all the methods.

\subsubsection{Methods for comparison}

\textbf{Single-task methods}
Since RV and FAZ segmentation are both dense prediction tasks, the state-of-the-art medical image or OCTA image segmentation methods were selected for comparison over both tasks, including U-Net~\cite{ronneberger2015u}, U-Net++~\cite{zhou2018unet++}, CE-Net~\cite{gu2019net},  OCTA-Net~\cite{ma2020rose}, TransUNet~\cite{chen2021transunet} and UTNet~\cite{gao2021utnet}.  {In addition, FARGO~\cite{peng2021fargo}, a multi-task method for RV and FAZ segmentation, was also selected.} For RVJ detection, we compared our model with two recent and representative methods: an RCNN-based method~\cite{Zhao2020VJ} and a multi-instance heatmap regression method~\cite{Hervella2020}, which are referred to as RB-Net  and HR-Net, respectively.

\textbf{Multi-task methods}
We also compared our method with four state-of-the-art MTL ones: UberNet~\cite{kokkinos2017ubernet}, Cross-stitch~\cite{Misra2016}, MTAN~\cite{liu2019end}, and MTI-Net~\cite{vandenhende2020mti}. For fair comparison, all these MTL methods employed the same task head, proposed in Section~\ref{sec:method}. 


\begin{table*}[]
        \setlength\tabcolsep{3pt}
        \renewcommand\arraystretch{1.1}
        \caption{{The detection results of RV, FAZ, and RVJ over different datasets using \textcolor{carminered}{\textbf{single-}} and  \textcolor{blue}{\textbf{multi-task}} learning approaches.}}%
        \centering
            \begin{tabular}{l|l|cc|cc|cc|cc}
            \hline \hline
            \multicolumn{1}{c}{\multirow{2}{*}{Dataset}}  & \multicolumn{1}{c}{\multirow{2}{*}{Methods}}  & \multicolumn{2}{c}{\textbf{RV segmentation}}  & \multicolumn{2}{c}{\textbf{FAZ segmentation}}    & \multicolumn{2}{c}{\textbf{RVJ detection}}       & \multicolumn{2}{c}{\textbf{RVJ classification}}  \\ \cline{3-10} 
            \multicolumn{1}{c}{}      &\multicolumn{1}{c}{}        & \multicolumn{1}{c}{DICE (\%)} & \multicolumn{1}{c}{BACC (\%)} & \multicolumn{1}{c}{DICE (\%)}  & \multicolumn{1}{c}{BACC (\%)}   & \multicolumn{1}{c}{RE (\%)}  & \multicolumn{1}{c}{F1-score (\%)} & \multicolumn{1}{c}{RE (\%)}   & \multicolumn{1}{c}{F1-score (\%)}  \\ \hline
            \multirow{10}{*}{\textbf{ROSE-O}}      & \textcolor{carminered}{U-Net}                & 70.32 $\pm$ 4.34             & 81.43 $\pm$ 3.65            & 93.82 $\pm$ 1.61         & 96.61 $\pm$ 1.57     & -----   & -----  & -----      & -----      \\  
                                      & \textcolor{carminered}{U-Net++}                    & 74.87 $\pm$ 4.03             &  83.48 $\pm$ 3.32    &   94.57 $\pm$ 2.16        &  {\textbf{97.85 $\pm$ 0.95}}    & -----   & -----  & -----      & -----    \\
                                      & \textcolor{carminered}{CE-Net}                     & 73.86 $\pm$ 3.65           &  82.64 $\pm$ 2.92    &   94.28 $\pm$ 2.35     &  97.77 $\pm$ 1.03     & -----   & -----  & -----      & -----   \\
                                      & \textcolor{carminered}{OCTA-Net}                   & 75.83 $\pm$ 4.45             &  84.24 $\pm$ 3.79    &   94.64 $\pm$ 2.43       &  97.59 $\pm$ 1.51  & -----   & -----  & -----      & -----    \\
                                      & \textcolor{carminered}{FARGO}                   & 74.59 $\pm$ 4.51             &  83.92 $\pm$ 3.29    &   94.42 $\pm$ 2.55       &  97.44 $\pm$ 1.48  & -----   & -----  & -----      & -----    \\
                                       & \textcolor{carminered}{TransUNet}                     & 74.90 $\pm$ 3.89           &  83.93 $\pm$ 4.07    &   94.33 $\pm$ 3.03     &  97.03 $\pm$ 2.19     & -----   & -----  & -----      & -----   \\
                                      & \textcolor{carminered}{UTNet}                   & 75.08 $\pm$ 2.95             &  84.33 $\pm$ 2.57    &   94.47 $\pm$ 1.96       &  97.36 $\pm$ 2.51  & -----   & -----  & -----      & -----    \\

                                      & \textcolor{carminered}{RB-Net}         & -----   & -----  & -----      & -----               & 53.43 $\pm$ 10.43                     & 62.86 $\pm$ 9.43            & 80.85 $\pm$ 5.15              & 67.13 $\pm$ 5.63           \\  
                                      &  \textcolor{carminered}{HR-Net}          & -----   & -----  & -----      & -----                    & 65.48 $\pm$ 8.32            &  70.17 $\pm$ 7.56    &   78.67 $\pm$ 4.78      &  66.43 $\pm$ 5.43     \\
                                 & \textcolor{blue}{UberNet}                     & 71.15 $\pm$ 4.55             &  80.57 $\pm$ 4.20    &   94.66 $\pm$ 2.12       &  96.63 $\pm$ 1.88    & 56.28 $\pm$ 5.61    &  64.74 $\pm$ 3.00    &   84.40 $\pm$ 3.92     &  67.83 $\pm$ 6.46  \\
                                 & \textcolor{blue}{Cross-Stitch}              & 74.48 $\pm$ 4.26             &  83.84 $\pm$ 4.36    &   94.57 $\pm$ 2.05      &  96.64 $\pm$ 1.38     &  {{73.67 $\pm$ 7.36}}      &  69.78 $\pm$ 3.68    &   79.11 $\pm$ 4.69      &  61.54 $\pm$ 7.86\\
                                     & \textcolor{blue}{MTAN}                       & 74.22 $\pm$ 4.74           &  82.81 $\pm$ 4.41    &   94.22 $\pm$ 2.11        &  96.21 $\pm$ 1.85    & 68.15 $\pm$ 10.05   &  67.67 $\pm$ 4.17    &   78.14 $\pm$ 5.51       &  62.70 $\pm$ 3.81     \\
                                       & \textcolor{blue}{MTI-Net}                   & 75.34 $\pm$ 4.73            &  83.63 $\pm$ 4.51    &   94.77 $\pm$ 1.92      &  96.43 $\pm$ 1.60   & 71.10 $\pm$ 10.21            &  68.74 $\pm$ 5.46    &   81.02 $\pm$ 3.58       &  63.74 $\pm$ 6.30  \\
                                      & \textcolor{blue}{VAFF-Net}                  & {\textbf{76.58 $\pm$ 4.74}}    &  {\textbf{84.62 $\pm$ 4.36}}    &  {\textbf{95.19 $\pm$ 1.57}}    &   {96.67 $\pm$ 1.43}     & {\textbf{73.88 $\pm$ 6.57}}            &   {\textbf{75.42 $\pm$ 3.91}}    &    {\textbf{84.26 $\pm$ 4.10}}      &   {\textbf{72.35 $\pm$ 4.29}}  \\ \hline
                                      
            \multirow{10}{*}{\textbf{ROSE-Z}}    & \textcolor{carminered}{U-Net}            & 78.39 $\pm$ 3.28           &  84.93 $\pm$ 3.77    &   91.17 $\pm$ 5.63         &  \textbf{96.98 $\pm$ 4.93}    & -----   & -----  & -----      & -----    \\
                                    & \textcolor{carminered}{U-Net++}                   & 80.12 $\pm$ 2.87            &  85.99 $\pm$ 2.34    &   92.53 $\pm$ 6.55        &  96.89 $\pm$ 5.41     & -----   & -----  & -----      & -----                    \\
                                      & \textcolor{carminered}{CE-Net}                    & 78.93 $\pm$ 4.36            &  85.66 $\pm$ 3.40    &   90.27 $\pm$ 6.35       &  98.81 $\pm$ 5.90   & -----   & -----  & -----      & -----                   \\
                                      & \textcolor{carminered}{OCTA-Net}                  & 80.44 $\pm$ 3.17           &  86.52 $\pm$ 2.49    &   92.05 $\pm$ 5.29       &  96.92 $\pm$ 6.45   & -----   & -----  & -----      & -----                  \\
                                      & \textcolor{carminered}{FARGO}                  & 78.97 $\pm$ 3.11          &  85.44 $\pm$ 3.71    &   92.20 $\pm$ 4.68        &  96.21 $\pm$ 5.45   & -----   & -----  & -----      & -----                  \\
                                      & \textcolor{carminered}{TransUNet}                     & 79.77 $\pm$ 2.97           &  86.35 $\pm$ 3.31   &   91.21 $\pm$ 2.35     &  96.04 $\pm$ 4.03     & -----   & -----  & -----      & -----   \\
                                      & \textcolor{carminered}{UTNet}                   & 80.40 $\pm$ 3.26             &  86.20 $\pm$ 2.28   &   91.96 $\pm$ 3.70       &  96.06 $\pm$ 4.06  & -----   & -----  & -----      & -----    \\
                                       & \textcolor{carminered}{RB-Net}           & -----   & -----  & -----      & -----    & {\textbf{75.34 $\pm$ 10.25}}      &  60.75 $\pm$ 9.29    &   74.99 $\pm$ 6.91    &  50.12 $\pm$ 7.05                \\
                                      & \textcolor{carminered}{HR-Net}        & -----   & -----  & -----      & -----       & 64.45 $\pm$ 10.09        &  64.90 $\pm$ 8.52    &   69.56 $\pm$ 6.59    &  54.87 $\pm$ 7.02      \\
                                      & \textcolor{blue}{UberNet}                & 77.14 $\pm$ 3.82            &  82.60 $\pm$ 3.71    &   90.39 $\pm$ 4.50        &  95.44 $\pm$ 2.91      & 58.48 $\pm$ 14.21   &  57.94 $\pm$ 13.11    &   76.93 $\pm$ 15.10    &  65.12 $\pm$ 16.19             \\
                                     & \textcolor{blue}{Cross-Stitch}             & 77.80 $\pm$ 3.42            &  83.98 $\pm$ 1.93    &   91.81 $\pm$ 5.96      &  95.69 $\pm$ 4.03        & 61.34 $\pm$ 11.59          &  59.22 $\pm$ 7.99    &   74.86 $\pm$ 11.28     &  61.44 $\pm$ 11.05             \\
                                     & \textcolor{blue}{MTAN}                      & 78.57 $\pm$ 3.16            &   {{85.29 $\pm$ 3.49}}    & 90.14 $\pm$ 8.81      &95.66 $\pm$ 4.49    & 61.28 $\pm$ 8.83    & 62.25 $\pm$ 8.62    & 73.68 $\pm$ 7.49     &  58.18 $\pm$ 8.43               \\
                                     
                                      & \textcolor{blue}{MTI-Net}                   & 79.72 $\pm$ 2.08           &  \textbf{86.71 $\pm$ 3.74}    &     91.84 $\pm$ 5.72      &  95.87 $\pm$ 3.20    & 57.62 $\pm$ 11.10    &  59.49 $\pm$ 7.32    &   69.34 $\pm$ 11.23       &  54.78 $\pm$ 11.05                \\
                                     & \textcolor{blue}{VAFF-Net}                  &  {\textbf{80.84 $\pm$ 2.93}}          &  86.05 $\pm$ 3.58    &    {\textbf{92.71 $\pm$ 3.90}}       &{{96.92 $\pm$ 3.07}}    & {63.75 $\pm$ 11.18}     &{\textbf{63.12 $\pm$ 8.65}}    & {\textbf{78.02 $\pm$ 4.55}}       &    {\textbf{67.75 $\pm$ 9.73}}                \\ \hline
            \multirow{10}{*}{\textbf{ROSE-H}}    & \textcolor{carminered}{U-Net}            & 78.60 $\pm$ 2.55           &  83.87 $\pm$ 1.76    &   74.28 $\pm$ 23.36      &  92.91 $\pm$ 9.88   & -----   & -----  & -----      & -----   \\
                                     & \textcolor{carminered}{U-Net++}                   & 78.74 $\pm$ 2.35           &  83.73 $\pm$ 1.52    &   76.25 $\pm$ 15.05        &  94.01 $\pm$ 6.33    & -----   & -----  & -----      & -----        \\
                                      & \textcolor{carminered}{CE-Net}                    & 80.07 $\pm$ 1.95           &  85.20 $\pm$ 1.43    &   80.42 $\pm$ 11.56        &  91.14 $\pm$ 8.42    & -----   & -----  & -----      & -----               \\
                                      & \textcolor{carminered}{OCTA-Net}                  & {80.53 $\pm$ 1.97}           &  {\textbf{85.73 $\pm$ 1.39}}    &   79.84 $\pm$ 10.73        &  93.02 $\pm$ 6.33   & -----   & -----  & -----      & -----        \\
                                      & \textcolor{carminered}{FARGO}                  & {79.17 $\pm$ 2.30}           &  {{84.24 $\pm$ 1.46}}    &   75.22 $\pm$ 16.35        &  94.80 $\pm$ 4.77   & -----   & -----  & -----      & -----        \\
                                      & \textcolor{carminered}{TransUNet}                     & 78.63 $\pm$ 2.01           &  84.06 $\pm$ 1.93    &   77.96 $\pm$ 10.43     &  93.62 $\pm$ 5.96    & -----   & -----  & -----      & -----   \\
                                      & \textcolor{carminered}{UTNet}                   & 79.34 $\pm$ 1.49             &  85.21 $\pm$ 2.50    &   78.44 $\pm$ 13.09       &  93.91 $\pm$ 7.18  & -----   & -----  & -----      & -----    \\
                                     & \textcolor{carminered}{RB-Net}            & -----   & -----  & -----      & -----      & {66.87 $\pm$ 12.92}    &  61.87 $\pm$ 5.83    &   83.71 $\pm$ 4.05      &  66.00 $\pm$ 5.88               \\
                                     & \textcolor{carminered}{HR-Net}             & -----   & -----  & -----      & -----    & 57.00 $\pm$ 12.91          &  64.82 $\pm$ 9.81    &   79.32 $\pm$ 4.29     &  68.22 $\pm$ 6.80  \\
                                    & \textcolor{blue}{UberNet}                & 76.23 $\pm$ 2.02      &  82.47 $\pm$ 1.71    &   67.75 $\pm$ 16.89       &  97.58 $\pm$ 1.54    & 71.60 $\pm$ 6.96    &  67.10 $\pm$ 3.61    &   82.13 $\pm$ 4.56   &  66.24 $\pm$ 7.43              \\
                                    & \textcolor{blue}{Cross-Stitch}             & 79.84 $\pm$ 2.60          &  83.93 $\pm$ 2.12    &   68.79 $\pm$ 16.91       &  97.54 $\pm$ 1.57    &   {\textbf{80.08 $\pm$ 7.23}}       &  69.44 $\pm$ 5.15    & 82.52 $\pm$ 4.45       &  65.28 $\pm$ 5.32                \\
                                     & \textcolor{blue}{MTAN}                      & 78.18 $\pm$ 3.06     &  82.77 $\pm$ 2.32    &   68.12 $\pm$ 15.08     &  97.71 $\pm$ 1.78     & 67.36 $\pm$ 6.04       &  65.81 $\pm$ 3.28    &   81.48 $\pm$ 4.24   &  65.91 $\pm$ 6.81                \\
                                     & \textcolor{blue}{MTI-Net}                   & 79.17 $\pm$ 2.33     &  83.79 $\pm$ 1.85    &   69.39 $\pm$ 14.12       &    {\textbf{98.20 $\pm$ 0.99}}    & 73.45 $\pm$ 8.42     &  69.52 $\pm$ 4.54    &   83.02 $\pm$ 3.85    &  68.89 $\pm$ 5.31           \\
                                     & \textcolor{blue}{VAFF-Net}                &{\textbf{81.62 $\pm$ 2.24}}      & {85.26 $\pm$ 1.69}    & {\textbf{81.06 $\pm$ 11.30} }     &  {94.73 $\pm$ 3.84}    & {{79.38 $\pm$ 10.20}}       &   {\textbf{71.17 $\pm$ 5.19}}    &    {\textbf{84.15 $\pm$ 4.01}}      &   {\textbf{72.71 $\pm$ 5.18}}   \\ \hline \hline
        \end{tabular}
        \label{tab:RV-FAZ}
        \end{table*}

 \begin{table*}[t]
    \renewcommand\arraystretch{1.1}
    \setlength\tabcolsep{4pt}
    \caption{{Effectiveness analysis of the VGM in VAFF-Net over the \textbf{ROSE-O} dataset. }}
    \centering
        \begin{tabular}{c|ll|ll|ll|ll}
        \hline \hline  \multicolumn{1}{c}{\multirow{2}{*}{{Method}}}  & \multicolumn{2}{|c}{{\textbf{RV segmentation}}}       & \multicolumn{2}{|c}{{\textbf{FAZ segmentation}}}  & \multicolumn{2}{|c}{{\textbf{RVJ detection}}}   & \multicolumn{2}{|c}{{\textbf{RVJ classification}}} \\ \cline{2-9} 
         \multicolumn{1}{c}{}     & \multicolumn{1}{|c}{DICE (\%)}  & \multicolumn{1}{c}{BACC (\%)} & \multicolumn{1}{|c}{DICE (\%)}  & \multicolumn{1}{c}{BACC (\%)} & \multicolumn{1}{|c}{RE (\%)}  & \multicolumn{1}{c}{F1-score (\%)}& \multicolumn{1}{|c}{RE (\%)}  & \multicolumn{1}{c}{F1-score (\%)} \\ \hline
         MAX              & 74.58 $\pm$ 4.39        &  83.46 $\pm$ 4.74    & 94.59 $\pm$ 1.96          & 96.63 $\pm$ 1.40   & 70.44 $\pm$ 6.36    & 67.17 $\pm$ 4.62     & 82.55 $\pm$ 3.77    & 60.45 $\pm$ 6.61 \\  
         MIN             &{73.90 $\pm$ 4.61}    & {82.85 $\pm$ 4.61}    & {94.28 $\pm$ 1.63}     & {95.86 $\pm$ 1.4}  & 65.10 $\pm$ 9.77    & 67.39 $\pm$ 5.25     & 82.13 $\pm$ 4.19    & 61.59$\pm$ 5.54  \\ 
         AVG            & 73.55 $\pm$ 4.68        &  82.51 $\pm$ 4.66    & 93.65 $\pm$ 2.40       & 96.48 $\pm$ 1.56  & 72.32 $\pm$ 8.33    & 69.99 $\pm$ 5.51      & 81.62 $\pm$ 3.34    & 61.22 $\pm$ 6.15  \\  
         SUM            &{71.96 $\pm$ 4.78}      & {81.72 $\pm$ 4.33}    & {94.76 $\pm$ 1.87}     & {96.65 $\pm$ 1.61}  & 65.28 $\pm$ 8.94   & 68.44 $\pm$ 6.17     & 82.05 $\pm$ 6.06    & 71.97 $\pm$ 6.95 \\ 
         VGM            &{\textbf{76.58 $\pm$ 4.74}}    & {\textbf{84.62 $\pm$ 4.36}}    & {\textbf{95.19 $\pm$ 1.57}}     & {\textbf{96.67 $\pm$ 1.43}}    & {\textbf{73.88 $\pm$ 6.57}}     &  {\textbf{75.42 $\pm$ 3.91}}     &   {\textbf{85.26 $\pm$ 4.10}}     &  {\textbf{{72.35 $\pm$ 4.29}}}     \\ 
         \hline \hline
    \end{tabular}
    \label{tab-ablation-VGM}
\end{table*}   
\subsubsection{Subjective comparisons}

Fig.~\ref{fig-resOCTA} shows the detection results of RAJ, RV and FAZ using different approaches over two randomly selected OCTA images from the ROSE-O and ROSE-H datasets respectively.
(Due to the page limit, we only provided one example result from each dataset for comparison). The top rows for ROSE-O and ROSE-H in Fig.~\ref{fig-resOCTA} show the RVJ detection and classification results. It may be observed that MTI-Net performs better in 
RVJ detection  when compared with MTAN and UberNet, as indicated in the representative patches. However, MTI-Net still suffers from the misclassification issue - it falsely identifies a bifurcation as a crossing 
(indicated by the red arrow). In contrast, our VAFF-Net method has shown its superiority in both the detection and classification tasks. 

The middle rows for ROSE-O and ROSE-H in Fig.~\ref{fig-resOCTA} illustrate the RV segmentation results. The benefit of the proposed method for RV segmentation may be observed from the representative regions (indicated by the red arrows). It may be seen that the competing methods achieved relatively poor performance in regions with low contrast - they preserve fine vessels poorly. In contrast, the proposed VAFF-Net yields more visually informative results, and is able to detect thinner vessels more completely.

The FAZ segmentation results are shown in the bottom rows for ROSE-O and ROSE-H in Fig.~\ref{fig-resOCTA}. Compared to other methods, it may be seen that our method displays less tendency to over- (green area) and under-segmentation (red area). That is because our method can make effective use of the DVC image via the VGM module: the DVC reveals the boundary of the FAZ more clearly than the other slabs, while it is difficult for the competing MTL methods to achieve this.

\subsubsection{Quantitative comparisons}
In order to demonstrate the superiority of the proposed method on different tasks, the quantitative results obtained by both single- and multi-task learning methods are shown in Table~\ref{tab:RV-FAZ}. 
Overall, the proposed VAFF-Net achieves promising RV and FAZ segmentation, and RVJ detection and classification results when compared with either single- or multi-task learning approaches. 

For example, our method gives the best RV segmentation performance over the ROSE-O dataset, and the second best over ROSE-H - only 0.47\% lower than OCTA-Net. This is because OCTA-Net was specifically designed for vessel segmentation in OCTA images, and the two-stage network is able to extract finer vessel structures. 
For the FAZ segmentation, our method achieves significantly higher performances over the  ROSE-O and ROSE-Z datasets, with the single exception that the accuracy is 1.18\% lower than that of U-Net++. 
For the RVJ detection and classification, our method achieves the best performance over all the three datasets, except on the ROSE-Z, where the RE of the RVJ detection is lower than that of RB-Net. This is because it sacrifices the accuracy of the RVJ classification for a higher detection accuracy. 
{On the ROSE-O and ROSE-Z datasets, the transformer achieved results comparable to CNNs-based methods. However, the transformer-based approach performed worse than the state-of-the-art CNN-based ones (e.g. OCTA-Net) on ROSE-H, which means that such methods are less capable of generalisation. This is because it has a large number of parameters, and training such a model with good generalization capabilities requires large-scale datasets.  
However, the datasets used in our task are relatively small and thus may lead it to overfitting.}

While noting that the ROSE-H dataset was used for testing only in our experiments, we also see that our method  produces higher performances in RV segmentation in DICE and RVJ classification in F1-score than those over ROSE-O and ROSE-Z. This may imply that the vascular presentation in ROSE-H (or in images acquired using the Heidelberg Spectralis OCT2 system) is more visible than in either ROSE-O or ROSE-Z.  

{It may be observed that there is a large discrepancy between DICE and BACC scores in Table~\ref{tab:RV-FAZ} for the FAZ segmentation over ROSE-H. 
This is caused by the domain gap and the relevant metrics, which are also considered as a critical issue in medical imaging. In this study, the images in ROSE-O  were captured using the Optovue Avanti RTVue XR, while ROSE-H was acquired using the Heidelberg Spectralis OCT2. It is very worth noting that in order to verify the generalizability  of our method, we used the former for training, but the latter for testing.  As shown in Fig.~\ref{fig-resOCTA}, large domain gap may be observed, especially in the FAZ area.
In this case, all the methods show relatively larger fluctuation in performing the FAZ segmentation. This means that the segmentation includes relatively more false positives (FPs) as over-segmentation and false negatives (FNs) as under-segmentation. 
{While TN is much larger than either FN or FP and TP is relatively small, these FPs and FNs influence DICE more significantly than BACC.}
}



\subsubsection{Comparison of the parameter size}
In addition, we compare the parameter size of our model against those of state-of-the-art MTL methods: UberNet (41.37M), Cross-stitch (65.50M), MTAN (44.79M), MTI-Net (94.29M), and VAFF-Net (34.73M). The number in the bracket indicates parameter size.  We can observe that our VAFF-Net has a much smaller size than all the other MTL methods. This is because our method not only shares parameters between the encoders, but the task-specific VGMs may select features for each task from the common ones,  avoiding the need to extract features for each task individually.

\begin{figure}[t]
    \centering{
    \includegraphics[width=0.9\linewidth]{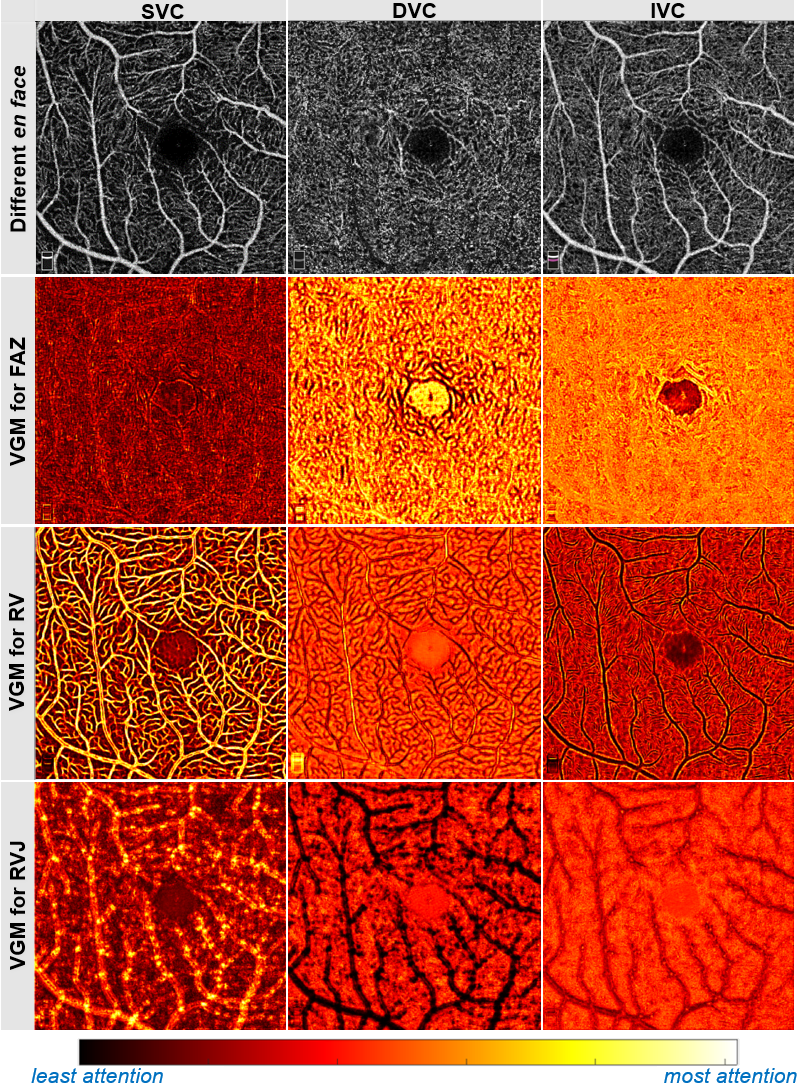}
    }
    \caption{Attention maps of the proposed VGM over SVC, DVC, and IVC in terms of different detection tasks.  }
       \label{fig-gate}
\end{figure}

\section{Discussions}
In this section, we carry out more experiments to demonstrate the importance of the proposed VGM and task head in  detection of multiple retinal structures. We also illustrate the detection performance of the proposed VAFF-Net over color fundus images to show its generalizability.

\subsection{Effectiveness of the VGM and its visualization}
The task-specific voting gate modules may adaptively learn how to perform feature fusion for different tasks. Specifically, the VGM learns weights according to the inputs for the selection of the features from different \textit{en face} images. To investigate the effectiveness of the VGM for each task, we replace the adaptive weighting of VGM with maximization (MAX), minimization (MIN), averaging (AVG) and summation (SUM) operations, respectively. Table~\ref{tab-ablation-VGM} shows the results over the ROSE-O dataset. It may be seen that adaptive weights of the VGM achieve the best results on all the tasks when compared with the other operations, by rather a large margin.

As explained initially in Sec.~\ref{sec:method}, our model can adaptively learn the VGMs in order to select feature maps obtained from different encoders for each task.  To better understand the mechanism of our approach, we show in Fig.\ref{fig-gate} the intermediate output (attention map) of the VGMs for each task in terms of different \textit{en face} angiograms. 
Brighter color (yellow) indicates increasing weight in performing feature selection, indicating the greater impact on the relevant task. 
For more accurate observation, the VGMs of different tasks pay different attentions to the importance of features from different encoders and spatial locations.
For example, in the FAZ segmentation task, the VGM pays more attention to the macular center region of the DVC, which is also in line with the clinical findings that the DVC image is better able to reveal the FAZ region~\cite{campbell2017detailed}.
In the case of the RV segmentation, the SVC reveals many more highlighted vessels than the DVC, as the SVC slab has particularly rich vascular structure.
The above findings indicate that the VGMs are able to select features adaptively for different tasks, allowing the model to share similar features among them while still taking into account the characteristics of each task.

\begin{table}[t]
    \setlength\tabcolsep{3pt}
    \renewcommand\arraystretch{1.12}
    \caption{{Performances of the proposed VAFF method with single- and multi-input over the \textbf{ROSE-O} dataset.}}
    \centering
        \begin{tabular}{c|ll|ll}
        \hline \hline  \multicolumn{1}{c}{\multirow{2}{*}{}}  & \multicolumn{2}{|c}{\textbf{RV segmentation}}       & \multicolumn{2}{|c}{\textbf{FAZ segmentation}}    \\ \cline{2-5} 
         \multicolumn{1}{c}{}     & \multicolumn{1}{|c}{DICE (\%)}  & \multicolumn{1}{c}{BACC (\%)} & \multicolumn{1}{|c}{DICE (\%)}  & \multicolumn{1}{c}{BACC (\%)} \\ \hline
        \textit{ Single-input}                              & 71.09 $\pm$ 4.74          &  81.16 $\pm$ 4.74            & 94.63 $\pm$ 2.58            & 95.78 $\pm$ 2.38           \\  
         \textit{Multi-input}              &{76.58 $\pm$ 4.74}    & {84.62 $\pm$ 4.36}    & {95.19 $\pm$ 1.57}     & {96.67 $\pm$ 1.43}    \\ \hline
         \multicolumn{1}{c}{\multirow{2}{*}{}}  & \multicolumn{2}{|c}{\textbf{RVJ detection}}       & \multicolumn{2}{|c}{\textbf{RVJ classification}}    \\ \cline{2-5} 
         \multicolumn{1}{c}{}     & \multicolumn{1}{|c}{RE (\%)} & \multicolumn{1}{c}{F1-score (\%)} & \multicolumn{1}{|c}{RE (\%)}  & \multicolumn{1}{c}{F1-score (\%)} \\ \hline
         \textit{Single-input}       &{72.47 $\pm$ 10.11}   & {70.01 $\pm$ 7.59}    & {83.72 $\pm$ 7.90}    & {70.95 $\pm$ 6.49}         \\
         \textit{Multi-input }    & 73.88 $\pm$ 6.57      &  {75.42 $\pm$ 3.91}   &  {85.26 $\pm$ 4.10}     & {72.35 $\pm$ 4.29}     \\
         \hline \hline
    \end{tabular}
    \label{tab:multi-input}
\end{table}


\begin{table*}[ht]
    \centering
    \setlength\tabcolsep{4pt}
        \renewcommand\arraystretch{1.12}
        \caption{{The results of RV segmentation and RVJ detection and classification using different \textcolor{carminered}{\textbf{single}-} and  \textcolor{blue}{\textbf{multi}-}task learning approaches over the \textbf{DRIVE} dataset. }}
            \begin{tabular}{lcccccc}
            \hline \hline
             \multicolumn{1}{l}{\multirow{2}{*}{Methods}}  & \multicolumn{2}{c}{\textbf{RV segmentation}}   & \multicolumn{2}{c}{\textbf{RVJ detection}}       & \multicolumn{2}{c}{\textbf{RVJ classification}}      \\ \cline{2-7} 
            \multicolumn{1}{c}{}       & \multicolumn{1}{c}{DICE (\%)}  & \multicolumn{1}{c}{BACC (\%)}  & \multicolumn{1}{c}{RE (\%)}  & \multicolumn{1}{c}{F1-score (\%)} & \multicolumn{1}{c}{RE (\%)}   & \multicolumn{1}{c}{F1-score (\%)}\\ \hline
                    \textcolor{carminered}{ U-Net}    & 77.80 $\pm$ 3.47         & 86.27 $\pm$ 3.93     & -----   & -----  & -----      & ----- \\  
                     \textcolor{carminered}{U-Net++}       & 78.47 $\pm$ 3.09            &  86.85 $\pm$ 3.60   & -----   & -----  & -----      & ----- \\
                     \textcolor{carminered}{CE-Net  }       & 79.03 $\pm$ 1.85            &  87.59 $\pm$ 2.84   & -----   & -----  & -----      & -----\\
                     \textcolor{carminered}{RB-Net  }  & -----   & -----  & 60.64 $\pm$ 9.97           &  56.36 $\pm$ 6.04    &   67.34 $\pm$ 4.42       &  69.90 $\pm$ 6.58               \\
                     \textcolor{carminered}{HR-Net   }  & -----   & -----  & {\textbf{70.32 $\pm$ 11.28}}           &  69.72 $\pm$ 9.88     &   77.08 $\pm$ 6.51       &  63.19 $\pm$ 7.59   \\
                     \textcolor{blue}{UberNet}              & 69.06 $\pm$ 2.85            &  80.68 $\pm$ 2.69     & 52.40 $\pm$ 10.58            &  56.02 $\pm$ 6.57    &   52.89 $\pm$ 5.92         &  48.15 $\pm$ 5.21  \\
                     \textcolor{blue}{Cross-Stitch}         & 78.29 $\pm$ 1.94           &  86.85 $\pm$ 2.37  & 57.48 $\pm$ 9.34            &  61.14 $\pm$ 4.58    &   67.36 $\pm$ 4.93        &  56.37 $\pm$ 6.26   \\
                     \textcolor{blue}{MTAN  }                         & 76.93 $\pm$ 1.90             &  76.93 $\pm$ 1.90     & 58.69 $\pm$ 8.10      &  63.46 $\pm$ 5.51    &   67.68 $\pm$ 5.33        &  61.46 $\pm$ 6.06   \\
                     \textcolor{blue}{MTI-Net}                       & 79.74 $\pm$ 1.60           &  88.26 $\pm$ 2.31       & 58.69 $\pm$ 8.10     &  58.64 $\pm$ 7.05    &   68.34 $\pm$ 5.30      &  58.64 $\pm$ 7.05  \\
                    \textcolor{blue}{ VAFF-Net}     &  {\textbf{81.09 $\pm$ 1.47}}     &  {\textbf{89.88 $\pm$ 2.35}}    &   {67.97 $\pm$ 9.05}         &   {\textbf{74.95 $\pm$ 4.68} }   &    {\textbf{77.39 $\pm$ 3.86}}        &  {\textbf{74.16 $\pm$ 4.37}} \\ \hline \hline
        \end{tabular}
        \vspace{-10pt}
        \label{tab:Drive-vessel}
\end{table*}






\subsection{Effectiveness of multiple \textit{en face} inputs}

All existing automated methods for extracting the multiple retina structures use the \textit{en face}  image of the IVC only~\cite{lin2021bsda,peng2021fargo,liang2021foveal}. We stated in Sec.~\ref{sec-intro} that a proper use of the depth information of different retinal layers in an OCTA image, i.e., IVC, SVC, and DVC, may conduce to a more accurate RV, RVJ and FAZ detection. 
In this subsection, we compare the detection performances when utilizing the IVC only, with multiple \textit{en face} as input: we define these two approaches as \textit{single-input}, and \textit{multi-input}, respectively.

Table~\ref{tab:multi-input} shows the performance of the proposed VAFF-Net with different inputs over the ROSE-O dataset. It may be observed that multi-input outperforms single-input in all the detection tasks, which demonstrates that the multi-input method can play an important role in multi-task learning for the extraction of the structures from OCTA images. For example, multi-input exhibits a large advantage over single-input by increases in DICE and ACC of about $5.5\%$ and $3.5\%$ on the vessel segmentation task, indicating that the use of the depth-resolved information is of benefit to the detection task.


\subsection{Performance on color fundus image}
\label{sec-general}
Although our VAFF-Net is designed for the analysis of OCTA images, it can also be used as a general multi-task learning tool for the analysis of images in different modalities. In the following experiment, the DRIVE dataset was employed. It includes 40 images with a resolution of $584 \times 565$ pixels. The groundtruth of vessel and vessel junctions are provided in~\cite{7493241}. We discuss the capability of our VAFF-Net in detection of RVs and RVJs from color fundus images. 

In this experiment, we set three identical color fundus images as the input of our model.  The VAFF-Net in this instance can be formulated as the model ensemble method~\cite{dong2020survey}.
The three encoders may be regarded as different subnetworks to obtain different features. The VGMs can adaptively weight and fuse the outputs of different subnetworks for different tasks.
In order to ensure the parameter diversity of different encoders, we adopt different parameter initialization strategies for the first convolutional block of different encoders, namely random initialization, Xavier initialization~\cite{glorot2010understanding}, and He initialization~\cite{he2015delving}. The voting module fuses the informative features for different tasks in an adaptive way.

Tables~\ref{tab:Drive-vessel} shows the quantification results of different methods for the RV segmentation, and RVJ detection and classification tasks. 
As may be seen, our model outperforms both single-task and other MTL methods in all the metrics on the RV segmentation, and RVJ detection and classification tasks, with the single exception that the recall of the RVJ detection is 2.35\% lower than that of HR-Net. This finding further reveals the capability of our VAFF-Net in detecting RV and RVJ from color fundus images in different modality.

\subsection{Effectiveness of junction head}
As aforementioned, the design of a task head for the RVJ detection task is one of our contributions. In order to demonstrate the superiority of the proposed junction head, we set the following experiment over the DRIVE dataset.

We first applied the encoder of ResNet50 as the backbone, and then compared the proposed junction head with two other state-of-the-art heads: RB-Net head~\cite{Zhao2020VJ} and HR-Net head~\cite{Hervella2020}. As may be seen in Table~\ref{tab:junchead}, RB-Net head performs better than HR-Net head in the RVJ classification, but worse than HR-Net head in the RVJ detection.
This is mainly due to the fact that RB-Net locates an RVJ by predicting the bounding-box, and thus has a lower localization accuracy, because the RVJ is a small target that covers only a few pixels in an OCTA image. 
In contrast, HR-Net distinguishes crossings from bifurcations by predicting a two-channel heatmap, allowing it to obtain a better detection performance compared with RB-Net (68.93\% vs. 55.88\% for F1-score). 
However, the number of the crossings is usually fewer than that of the bifurcations in OCTA images, which makes the model tend to under predict crossings, resulting in worse classification performance (62.82\% vs. 68.98\% for F1-score). This is because the error of falsely predicting a crossing as a bifurcation will be higher than the error when no RVJ is predicted at all.
Since our method combines the advantages of the other two methods, it achieves the best performance in both the detection and classification of RVJs.

\begin{table}[t]
    \setlength\tabcolsep{3pt}
    \renewcommand\arraystretch{1.12}
    \caption{The performance of different methods for RVJ detection and classification task over the \textbf{DRIVE} dataset.}
    \centering
        \begin{tabular}{lllll}
        \hline \hline
           \multicolumn{1}{c}{\multirow{2}{*}{Head}}  & \multicolumn{2}{c}{\textbf{RVJ detection}}       & \multicolumn{2}{c}{\textbf{RVJ classification}}    \\ \cline{2-5} 
         \multicolumn{1}{c}{}     & \multicolumn{1}{c}{RE (\%)}  & \multicolumn{1}{c}{F1-score (\%)} & \multicolumn{1}{c}{RE (\%)}    & \multicolumn{1}{c}{F1-score (\%)} \\ \hline
           RB-Net head   & 60.64 $\pm$ 9.97      &  55.88 $\pm$ 5.98    &   65.32 $\pm$ 4.32      &  68.98 $\pm$ 6.11        \\
           HR-Net head   & 68.98 $\pm$ 11.28        &  68.93 $\pm$ 9.88    &   75.08 $\pm$ 5.57       &  62.82 $\pm$ 6.68     \\
           Our head    & {\textbf{71.70 $\pm$ 7.75}}    &  {\textbf{76.70 $\pm$ 3.75}}    &  {\textbf{ 76.29 $\pm$ 4.84}}      &  {\textbf{72.73 $\pm$ 5.49}}        \\ \hline \hline
    \end{tabular}
    \label{tab:junchead}
    \end{table}




\section{Conclusion}

In this work, we have proposed a novel method for detection of retinal structures in OCTA images. Many studies have demonstrated that the quantification of the retinal structures obtained in OCTA images plays a vital role in the quantitative study and clinical decision-making of ophthalmopathological and neurodegenerative diseases.  To this end, we  proposed an end-to-end multi-task learning method for the joint segmentation, detection and classification of retinal vessels, foveal avascular zone  and retinal vascular junctions in OCTA images. The method exploits the layered structure information to promote each individual task. 
We argue that these tasks share some similarities: we therefore first employed three encoders that share weights (except the first convolution layer) to extract relevant features from different inputs.
However, since the characteristics of each task also differ in some aspects, we designed the task-specific voting gate modules to adaptively choose favored features from the different encoders. This voting-based feature integration strategy can automatically select features at two levels according to respective task requirements: 1) features from different layer \textit{en face} images; and 2) features at different spatial locations from an encoder, which is also important for each task.

In addition to the new detection method, we further constructed three OCTA datasets for the detection task of multiple retinal structures, and the experimental results show that our VAFF-Net outperforms on the whole both the state-of-the-art single-purpose methods and earlier multi-task learning methods. In addition, we also demonstrated that our model can be used as a general  multi-task learning tool for the analysis of images in other modalities, which was validated over a color fundus image dataset.
In future, we will extract the biomarkers from the segmented retinal vessels and FAZ and the detected RVJs in the given OCTA images and associate them with relevant eye-related diseases.  It is also our intention to perform accurate 3D
vessel reconstruction and feature quantification directly from 3D OCTA volumetric data in future.

\bibliographystyle{IEEEtran}
\bibliography{refs-short}

\end{document}


\title{\huge Supplemental Materials: Retinal Structure Detection in OCTA Image via Voting-based Multi-task Learning}


\maketitle

\section{More on subjective comparisons}
\vspace{-5pt}
Fig~\ref{fig-octa-sup} shows more qualitative results of different multi-task learning approaches over the {ROSE-H} and {ROSE-Z} datasets.
As can been seen from the results of the RVJ task in the first rows, our VAFF-Net method has shown its superiority in both detection and classification results that are are the closest to the ground truth. 
For the RV segmentation results in the second rows, compared with the competitors, the proposed VAFF-Net yields finer results,  it is able to detect the thinner and more continuous vessels (shown in red arrows).
The third rows of Fig~\ref{fig-octa-sup} illustrates the FAZ segmentation results. Our method has less over-segmentation (green area) and under-segmentation (red area) performances compared to other methods. However, it may be observed in the case from ROSE-H that all the methods has poorer FAZ segmentation in OCTA-H dataset when compared with the results of ROSE dataset, that is because we used the ROSE-trained model to test on the OCTA-H dataset. As these two datasets were  acquired by two different imaging devices, and it may appear some domain differences.

We also present the qualitative comparison between MTI-Net and VEFF-Net on the DRIVE dataset.  Fig.~\ref{fig-dirve} shows the RV segmentation and RVJ detection results, as can be seen, the proposed VAFF-Net still achieves satisfactory results when used as a general multi-task learning method. Compared with MTI-Net, VAFF-Net is significantly better at RV and RVJ detection in low contrast regions, and its results are closer to the ground truth.

\section{More experiments on discussion}
\subsection{Effectiveness of multiple \textit{en face} inputs}
To further validate the findings in the main manuscript Section VII.(B), we compared the detection performance on ROSE-Z and ROSE-H datasets when applying IVC and multiple \textit{en face} as inputs, which we defined as single- and multi-inputs, respectively. Table~\ref{tab-multi-more} shows that multiple inputs outperformed single inputs for all detection tasks in both datasets, indicating that multiple inputs play an important role in extracting structure in OCTA images.

\begin{table}[H]
        \renewcommand\arraystretch{1.1}
        \setlength\tabcolsep{4pt}
        \centering
        \caption{{Performances of single- and multi-input over the \textbf{ROSE-Z} and \textbf{ROSE-H} dataset. }}
        \vspace{-5pt}
            \begin{tabular}{c|c|ll|ll|ll|ll}
            \hline \hline
             \multicolumn{1}{c|}{\multirow{2}{*}{Dataset}}  & \multicolumn{1}{c}{\multirow{2}{*}{{Method}}}      & \multicolumn{2}{|c}{{\textbf{RV segmentation}}}       & \multicolumn{2}{|c}{{\textbf{FAZ segmentation}}}  & \multicolumn{2}{|c}{{\textbf{RVJ detection}}}   & \multicolumn{2}{|c}{{\textbf{RVJ classification}}} \\ \cline{3-10} 
            \multicolumn{1}{c|}{}   &\multicolumn{1}{c}{}     & \multicolumn{1}{|c}{DICE (\%)}  & \multicolumn{1}{c}{BACC (\%)} & \multicolumn{1}{|c}{DICE (\%)}  & \multicolumn{1}{c}{BACC (\%)} & \multicolumn{1}{|c}{RE (\%)}  & \multicolumn{1}{c}{F1-score (\%)}& \multicolumn{1}{|c}{RE (\%)}  & \multicolumn{1}{c}{F1-score (\%)} \\ \hline
            \multirow{2}{*}{ROSE-Z}  & Single-input      & 78.64 $\pm$ 1.94        &  86.07 $\pm$ 1.77    & 92.18 $\pm$ 8.06          & 95.37 $\pm$ 4.76   & 62.15 $\pm$ 10.16    & 61.05 $\pm$ 11.92     & 78.54 $\pm$ 11.21    & 67.72 $\pm$ 10.20 \\  
                                    & Multi-input      & {{80.58 $\pm$ 2.52}}          &  86.27 $\pm$ 2.15    &   {{92.40 $\pm$ 5.95}}       &{{96.64 $\pm$ 4.23}}    & {64.03 $\pm$ 10.08}     &{{62.72 $\pm$ 6.56}}    &{{79.72 $\pm$ 5.35}}       &   {{69.45 $\pm$ 8.27}}   \\ 
             \hline 
             \multirow{2}{*}{ROSE-H}  & Single-input    & 78.86 $\pm$ 2.33        &  82.42 $\pm$ 1.61    & 80.06 $\pm$ 12.08          & 94.80 $\pm$ 2.74   & 70.83 $\pm$ 11.70    & 70.92 $\pm$ 5.9     &{82.93 $\pm$ 3.50}     &   {70.07 $\pm$ 4.93}\\  
                                    & Multi-input  &{{81.62 $\pm$ 2.24}}      & {85.26 $\pm$ 1.69}    & {{81.06 $\pm$ 11.30} }     &  {94.73 $\pm$ 3.84}    & {{79.38 $\pm$ 10.20}}       &   {{71.17 $\pm$ 5.19}}    &    {{84.15 $\pm$ 4.01}}      &   {{72.71 $\pm$ 5.18}}   \\ 
             \hline \hline
        \end{tabular}
        \label{tab-multi-more}
    \end{table} 

\begin{figure}[t]
    \centering{
    \includegraphics[width=0.9\linewidth]{./figure/res-drive.png}
    }
    \vspace{-5pt}
    \caption{Qualitative results of our VAFF-Net and MTI-Net on the DRIVE dataset.}
    \label{fig-dirve}
    \vspace{-10pt}
\end{figure}

    \begin{table*}[h]
        \renewcommand\arraystretch{1.1}
        \setlength\tabcolsep{4pt}
        \caption{{Effectiveness analysis of the VGM in VAFF-Net over the \textbf{ROSE-Z} \textbf{ROSE-H} dataset. }}
        \centering
            \begin{tabular}{c|c|ll|ll|ll|ll}
            \hline \hline
             \multicolumn{1}{c|}{\multirow{2}{*}{Dataset}}  & \multicolumn{1}{c}{\multirow{2}{*}{{Method}}}      & \multicolumn{2}{|c}{{\textbf{RV segmentation}}}       & \multicolumn{2}{|c}{{\textbf{FAZ segmentation}}}  & \multicolumn{2}{|c}{{\textbf{RVJ detection}}}   & \multicolumn{2}{|c}{{\textbf{RVJ classification}}} \\ \cline{3-10} 
            \multicolumn{1}{c|}{}   &\multicolumn{1}{c}{}     & \multicolumn{1}{|c}{DICE (\%)}  & \multicolumn{1}{c}{BACC (\%)} & \multicolumn{1}{|c}{DICE (\%)}  & \multicolumn{1}{c}{BACC (\%)} & \multicolumn{1}{|c}{RE (\%)}  & \multicolumn{1}{c}{F1-score (\%)}& \multicolumn{1}{|c}{RE (\%)}  & \multicolumn{1}{c}{F1-score (\%)} \\ \hline
            \multirow{5}{*}{ROSE-Z}  & MAX   & 79.81 $\pm$ 1.75        &  85.81 $\pm$ 1.75    & 91.32 $\pm$ 6.81         & 95.56 $\pm$ 4.54   & 61.42 $\pm$ 5.21    & 61.05 $\pm$ 7.62     & 77.17 $\pm$ 8.83    & 66.42 $\pm$ 7.70 \\  
                                    & MIN      &{79.38 $\pm$ 1.89}    & {85.30 $\pm$ 1.83}    & {91.29 $\pm$ 1.63}     & {96.35 $\pm$ 1.4}  & 60.91 $\pm$ 9.09    & 61.13 $\pm$ 6.27     & 76.27 $\pm$ 1.69    & 65.23$\pm$ 10.55  \\ 
                                    & AVG            & 79.06 $\pm$ 1.96        &  85.52 $\pm$ 1.86   & 90.73 $\pm$ 5.22       & 96.54 $\pm$ 5.13  & 60.46 $\pm$ 10.90    & 60.18 $\pm$ 8.54      & 78.25 $\pm$ 7.16    & 67.27 $\pm$ 7.68  \\  
                                    & SUM            &{76.68 $\pm$ 2.25}      & {84.21 $\pm$ 2.00}    & {89.87 $\pm$ 7.03}     & {95.26 $\pm$ 3.57}  & 59.15 $\pm$ 11.16   & 60.05 $\pm$ 11.92     & 75.43 $\pm$ 11.24    & 66.25 $\pm$ 10.20 \\ 
                                    & VGM           & {\textbf{80.58 $\pm$ 2.52}}          &  \textbf{86.27 $\pm$ 2.15}    &   {\textbf{92.40 $\pm$ 5.95}}       &{\textbf{96.64 $\pm$ 4.23}}    & \textbf{64.03 $\pm$ 10.08}     &{\textbf{62.72 $\pm$ 6.56}}    &{\textbf{79.72 $\pm$ 5.35}}       &   {\textbf{69.45 $\pm$ 8.27}}    \\ 
             \hline 
             \multirow{5}{*}{ROSE-H}  & MAX              & 79.03 $\pm$ 2.43        &  83.01 $\pm$ 1.71  & 79.27$\pm$ 12.12         & \textbf{94.93$\pm$ 2.84}   & 72.10 $\pm$ 12.27    & 70.34 $\pm$ 6.33    & 82.30 $\pm$ 3.57    & 68.60 $\pm$ 5.05 \\  
                                    & MIN             &{79.44 $\pm$ 2.42}    & {83.33 $\pm$ 1.71}    & {78.73 $\pm$ 12.14}     & {94.54 $\pm$ 2.56}  & 71.99 $\pm$ 11.42    & 70.63 $\pm$ 5.76     & 82.45 $\pm$ 5.50    & 69.83$\pm$ 5.54 \\ 
                                    & AVG            & 79.48 $\pm$ 2.44        &  83.32 $\pm$ 1.74    & 77.28 $\pm$ 16.31       & 94.18 $\pm$ 4.18  & 74.70 $\pm$ 10.50    & 72.28 $\pm$ 4.07      & 81.48 $\pm$ 4.50    & 70.52 $\pm$ 6.43  \\  
                                    & SUM            &{78.30 $\pm$ 2.33}      & {82.42 $\pm$ 1.61}    & {77.86 $\pm$ 13.91}     & {94.87 $\pm$ 2.85}  & 76.76 $\pm$ 10.39   & \textbf{73.19 $\pm$ 4.85}     & 82.98 $\pm$ 3.74   & 69.86 $\pm$ 5.78\\ 
                                    & VGM            &{\textbf{81.62 $\pm$ 2.24}}      & \textbf{85.26 $\pm$ 1.69}    & {\textbf{81.06 $\pm$ 11.30} }     &  {94.73 $\pm$ 3.84}    & {\textbf{79.38 $\pm$ 10.20}}       &   {{71.17 $\pm$ 5.19}}    &    {\textbf{84.15 $\pm$ 4.01}}      &   {\textbf{72.71 $\pm$ 5.18}}    \\
             \hline \hline
        \end{tabular}
        \label{tab-VGM-more}
    \end{table*}

\subsection{Effectiveness of VGM}
In the main manuscript, we investigate the effectiveness of VGM using the ROSE-O dataset by replacing the adaptive weighting of VGM with maximization (MAX), minimization (MIN), averaging (AVG) and summation (SUM) operations, respectively.
Here,  as Table~\ref{tab-VGM-more} shown, we report more extensive experimental results on the ROSE-Z and ROSE-H datasets. As can be seen, the VGM clearly has a significant advantage over the other operations due to the adaptive weighting strategy.

\begin{figure*}[h]
    \centering{
    \includegraphics[width=0.95\linewidth]{./figure/octa-sup.png}
    }
    \vspace{-6pt}
    \caption{Visual comparison of the detection results of our method and different MTL methods on a randomly selected case from \textbf{ROSE-H} and \textbf{ROSE-Z}. For one case from top to bottom are: the results of RVJ detection and classification, RV segmentation, and FAZ segmentation, respectively. 
    }
    \label{fig-octa-sup}
    \vspace{-10pt}
\end{figure*}